\documentclass[journal=jacsat,manuscript=article]{achemso}

\usepackage[version=3]{mhchem} 
\usepackage{multirow,rotating}
\usepackage{xcolor}
\usepackage{array}

\setkeys{acs}{articletitle = true}

\author{Joan Enrique-Romero}
\affiliation[Universit\'e Grenoble Alpes]
{Univ. Grenoble Alpes, Institut de Plan\'etologie et Astrophysique de Grenoble (IPAG), 38000 Grenoble, France}
\alsoaffiliation[Universitat Aut{\`o}noma de Barcelona]
{Departament de Qu{\'i}mica, Universitat Aut{\`o}noma de Barcelona, 08193, Catalunya, Spain}
\author{Albert Rimola}
\affiliation[Universitat Aut{\`o}noma de Barcelona]
{Departament de Qu{\'i}mica, Universitat Aut{\`o}noma de Barcelona, 08193, Catalunya, Spain}
\email{albert.rimola@uab.cat}
\author{Cecilia Ceccarelli}
\affiliation[Universit\'e Grenoble Alpes]
{Univ. Grenoble Alpes, Institut de Plan\'etologie et Astrophysique de Grenoble (IPAG), 38000 Grenoble, France}
\email{cecilia.ceccarelli@univ-grenoble-alpes.fr}
\author{Piero Ugliengo}
\affiliation[Universit{\`a} degli Studi di Torino]{Dipartamento di Chimica and Nanostructured Interfaces and Surfaces (NIS), Universit{\`a} degli Studi di Torino, Via P. Giuria 7, 10125 Torino, Italy}
\author{Nadia Balucani}
\affiliation[Universit{\`a} degli Studi di Perugia]{Dipartamento di Chimica, Biologia e Biotecnologie, Universit{\`a} degli Studi di Perugia, Via Elce di Sotto 8, I-06123 Perugia, Italy}
\alsoaffiliation[INAF]{Osservatorio Astrofisico di Arcetri, Largo E. Fermi 5, 50125 Firenze, Italy}
\alsoaffiliation[Universit\`eGrenoble Alpes]{Univ. Grenoble Alpes, Institut de Plan\`etologie et Astrophysique de Grenoble (IPAG), 38000 Grenoble, France}
\author{Dimitrios Skouteris}
\affiliation[Master-Up, Perugia, Italy]{Master-Up, Strada Vicinale Sperandio 15, I-06123 Perugia Italia Perugia, Italy}

\title[]
  {Reactivity of HCO with CH$_3$ and NH$_2$ on Water Ice Surfaces. A Comprehensive Accurate Quantum Chemistry Study}

\setkeys{acs}{keywords = true}
\abbreviations{ISM,\colorbox{yellow}{ASW}}

\begin{document}








\begin{abstract}
  Interstellar complex organic molecules (iCOMs) can be loosely
    defined as chemical compounds with at least six atoms in which
  at least one is carbon. The observations of iCOMs in
    star-forming regions have shown that they contain an important
    fraction of carbon in a molecular form, which can be used to
    synthesize more complex, even biotic molecules. Hence, iCOMs are
    major actors in the increasing molecular complexity in space and they might have played a role in the origin of terrestrial life. Understanding
  how iCOMs are formed is relevant for predicting the ultimate organic chemistry
  reached in the interstellar medium. One possibility is that
    they are synthesized on the interstellar grain icy surfaces, via
    recombination of previously formed radicals. The present work
  focuses on the reactivity of HCO with CH$_3$/NH$_2$ on the grain icy
  surfaces, investigated by means of quantum chemical
  simulations. The goal is to carry out a systematic study using
    different computational approaches and models for the icy
    surfaces. Specifically, DFT computations have been bench-marked with CASPT2 and CCSD(T) methods, and the ice
    mantles have been mimicked with cluster models of 1, 2, 18 and 33
    H$_2$O molecules, in which different reaction sites have been
    considered. Our results indicate that the HCO +
    CH$_3$/NH$_2$ reactions, if they actually occur, have two major competitive channels:
    the formation of iCOMs CH$_3$CHO/NH$_2$CHO, or the formation of CO +
    CH$_4$/NH$_3$. These two channels are either barrierless or present
    relatively low ($\leq 10$ kJ/mol equal to about 1200 K) energy
    barriers. Finally, we briefly discuss the astrophysical
    implications of these findings.
\end{abstract}

\section{1. Introduction}
It has been long demonstrated that star forming regions are places with a
rich organic chemistry (e.g. \cite{rubin1971, cazaux2003, herbst2009, caselli2012, belloche2017, mcguire2018}).
Although there are no proofs that organic molecules formed in the
interstellar medium (ISM) did play a role in the emergence of
terrestrial life, there is mounting evidence that they were inherited
by the small objects of the Solar System: for example, carbonaceous
chondrites and comets contain a wide variety of organic molecules,
some of them probably being a direct heritage of the ISM based on
their relative abundances and ratio of deuterated versus hydrogenated
species (e.g. \cite{bockelee2011, caselli2012, ceccarelli2014,
  bianchi2018}).

Knowing how detected interstellar complex organic molecules (iCOMs:
C-bearing molecules with at least six atoms \cite{ceccarelli2017})
are formed and destroyed is not only important per se but also because
it is the only way to understand the ultimate organic complexity
present in the ISM. Indeed, there is an intrinsic limit to the
detection of large iCOMs (excluding linear chains), which is caused by
the fact that the larger the molecule the larger the number of
rotational transitions (because of the larger number of functional
groups of the iCOM) and, consequently, the weaker the intensity of the
lines. As a result, the numerous and weak lines of large iCOMs produce
a ``grass'' of lines in the spectra, which makes the identification of
a large molecule eventually impossible. Therefore, there is a limit to
the largest detectable iCOM, and this has a direct consequence: we
need to rely on our knowledge of the processes to predict which large
molecules are synthesized in the ISM.

How iCOMs are formed is a question that has baffled astrochemists for
decades. In the 90s it was thought that gas-phase reactions were the
dominant formation processes (e.g. \cite{charnley1992, caselli1993}).
However, subsequent astronomical observations (e.g. \cite{ceccarelli2001, cazaux2003}),
laboratory experiments (e.g. \cite{geppert2005}),
and theoretical calculations (e.g. \cite{horn2004})
challenged this synthetic route model. As a result, a new paradigm was
proposed, which assumes that most (if not all) iCOMs are formed on the
surfaces of icy interstellar grains.\cite{garrod2006, herbst2017}
According to this paradigm, radicals are created by the UV photons
impinging on the ice grain mantles, which are formed during the cold
cloud/prestellar phase and contain simple hydrogenated species (mostly
water but also other species, such as methanol and ammonia, in smaller
quantities). With the warming up of the environment caused by the
birth of a protostar, the radicals trapped in the ice can diffuse on
the grain surfaces and react between them to form iCOMs. The reaction
process is assumed to be a direct combination of radicals, which are
considered as ``lego-like" blocks.

This paradigm is usually assumed in current gas-grain astrochemical
models (e.g. \cite{garrod2008, acharyya2015, ruaud2015modelling,
  drozdovskaya2016, vasyunin2017}).
However, there are still many uncertainties that makes the paradigm
not fully validated. First, it is not clear that the lego-like radical
combination is actually an effective process (refs:
e.g. \cite{rimola2018, enrique2016}).
Second, it appears that the role of gas-phase reactions has been
underestimated in the past.
\cite{balucani2015,Spezia2016Formamide,Taquet2016PSO,Suzuki2016_glycine,Codella2017Solis, bianchi2018,skouteris2018interstellar,rosi2018formation,skouteris2018genealogical}
Finally, models based on the ``exclusive grain-surface"paradigm are
unable to reproduce the observed abundance of several iCOMs
(e.g. \cite{Coutens2016, Muller2016Exploring, Ligterink2018}).

The numerous laboratory experiments reported in the literature are extremely useful to study the possible processes occurring on ices illuminated
by UV photons and/or irradiated by energetic particles
  (e.g. \cite{STRAZZULLA2001,Palumbo2008, Woods_2013, Fedoseev2014, Fedoseev2015, Linnartz2015PhysChemRev, deBarros2016,  Butscher2017, Fedoseev2017, Arumainayagam2019, Dulieu2019, PEREIRA2019}). 
  However, investigating
surface-induced iCOMs formation by means of experimental techniques is
extremely difficult, if not impossible, since the actual interstellar
conditions cannot exactly be reproduced in terrestrial
laboratories. For example, the UV irradiation used in laboratory
experiments is, for practical reasons, more than million times larger
than the one impinging on the interstellar grain mantles: this causes
an instantaneous injection of energy of several orders of magnitude
larger than that absorbed by the interstellar grain mantles and,
consequently, likely introduces different reaction routes
(e.g. energy barriers are probably overcome in laboratory experiments
while they are not in the cold ISM).  Similarly, the H-flux in
experiments is also extremely larger compared with the actual ISM
conditions, causing a dramatic amplification of the results towards a
full saturation of the compounds. A suitable alternative, and
certainly a complementary method for such investigations, is the use of
theoretical simulations based on quantum chemical
calculations. Indeed, these calculations provide a description of the
surface reactions from an atomistic perspective, providing unique,
relevant energetic information, and, accordingly, they can be useful
to assess the validity of the ``exclusive grain-surface" paradigm.

In this work, we focus on two reactions occurring on amorphous solid
water (ASW) surfaces: HCO + CH$_3$ and HCO + NH$_2$. In the
``exclusive grain-surface" paradigm mentioned above, it is expected
that the radical coupling produces two iCOMs: CH$_3$CHO
(acetaldehyde) and NH$_2$CHO (formamide). However,
previous works by some of us \cite{enrique2016, rimola2018}
showed that other reactive channels can compete with the iCOM
formation. Specifically, the two reactions can lead to the formation
of CO + CH$_4$ and CO + NH$_3$, respectively, in which the H atom of
HCO is transferred to the radical partner. Similar processes were
identified computationally when HCO reacts with CH$_3$ on surfaces of
CO-pure ices.\cite{kastner2019}

The goal of the present work is to carry out a systematic study of two
  reactions, i.e.,  HCO + CH$_3$ and HCO + NH$_2$, considered here as prototype reactions for the formation of iCOMs, using different
  approximations for the calculations and models of ASW with the aim to: (1) understand how the different methods and models affect the
  results, (2) individuate the most convenient methods and models to use in future
  calculations of other similar radical-radical systems, and (3) identify are the products of the reactions for different conditions.  To this
  end, the present work focuses on the following three points:
\begin{enumerate}
\item {\it Methodology benchmark:} (i) The energy barriers for
  reactions between the two couples of radicals in the presence of 1 and 2
  water molecules are computed with two DFT methods (B3LYP and BHLYP)
  and compared to the values calculated with the multi-reference
  CASPT2 method; (ii) The interaction of the three radicals (CH$_3$,
  HCO and NH$_2$) with 1 and 2 water molecules is studied and bench-marked
  taking as reference the binding energies computed at the CCSD(T)
  level.
\item {\it Radical-surface binding enthalpies:} We study the
  binding of the three radicals (CH$_3$, HCO and NH$_2$) to an ASW
  cluster model of 18 water molecules and to a
  larger cluster of 33 water molecules sporting two different
  morphological sites, a cavity and its side.
\item {\it Radical-radical reactivity:} The reactivity of the two
  radical couples is studied (i) on the 18 water molecules cluster, and (ii) on the two different
  morphological sites of the 33 water molecules cluster.
\end{enumerate}

It is worth mentioning that the systems to deal with here sport an
additional complexity from an electronic structure point of view. That
is, two radicals interacting with water ice, in which the unpaired
electrons have opposite signs, constitute a singlet biradical
system. Describing this electronic situation with quantum chemistry
calculations is delicate. Ab initio multi-reference methods are a good
choice to describe biradical systems but they are extremely expensive
and, accordingly, unpractical for large systems. Alternatively, a good
compromise between accuracy and computational cost is the DFT
broken-symmetry approach.\cite{abe2013}. By using this method,
however, our own experience\cite{enrique2016, rimola2018}
indicates that one has to be sure that the initial guess wave function
corresponds to the actual singlet biradical state (i.e., the unpaired
electrons being localized on the corresponding radicals and with
opposite spin signs), as it may well happen that the initial wave
function represents wrongly a closed-shell-like situation, in which
the unpaired electrons are 50\% distributed among the two
radicals. Results derived from one or the other situation are
dramatically different.

This article is organized as follows. First the adopted methods are
presented (\S\, 2), then the results are provided following 
the 3 objectives described above (\S\, 3) and finally
a discussion, including the astrophysical implications, and the
conclusions are presented in \S\, 4 and 5, respectively.

\section{2. Methods}

All DFT and CCSD(T) calculations were performed using the
GAUSSIAN09\cite{gaussian09} software package, while the
multi-reference calculations were carried out with the OpenMolcas
18.09\cite{Molcas,Molcas7,Molcas8,OpenMolcas} program.

Stationary points of the potential energy surfaces (PESs) were fully
optimized using two hybrid density functional theory (DFT) methods:
B3LYP and BHLYP.  These methods have the same Lee, Yang, Parr
correlation functional (LYP)\cite{lee1988} but differ on the exchange
functionals: the Becke's three parameter (B3), which includes a 20\%
of exact exchange in its definition,\cite{b3lyp-becke1993} and the
Becke's half-and-half (BH), which mixes the pure DFT and the exact
exchange energy in a 1:1 ratio.\cite{bhandhlyp-becke1993}. For B3LYP calculations, the Grimme's D3 dispersion term\cite{D3-grimme2010} was accounted for during the geometry optimizations. In contrast, for BHLYP, both the D2\cite{D2-grimme2006} and D3 dispersion terms were included in a posteriori way onto the pure BHLYP optimized geometries.

A calibration study was first
carried out for (i) the NH$_2$ + HCO and CH$_3$ + HCO reactivities
in the presence of 1 and 2 water molecules (W1 and W2,
respectively) and (ii) the interaction of each radical with W1 and
W2.  For this calibration study, the DFT methods were combined
with the Pople's 6-311++G(2df,2pd) basis set. As reference for the
reactivity results we used single point energy calculations at the
multi-reference CASPT2 level combined with the Dunnning's cc-pVTZ
basis using as initial guess the orbitals generated at CASSCF(2,2)
level.  
In the same way, single
point energy calculations at the CCSD(T)/aug-cc-pVTZ level were also carried out in
order to compare them to CASPT2 values. Regarding the interaction
energies, the same DFT methods were compared to
CCSD(T)/aug-cc-pVTZ level. All the single point energy calculations for this benchmark
study were carried out on the B3LYP-D3 optimized geometries.

\begin{figure}[H]
    \centering
    \includegraphics[width=0.98\textwidth]{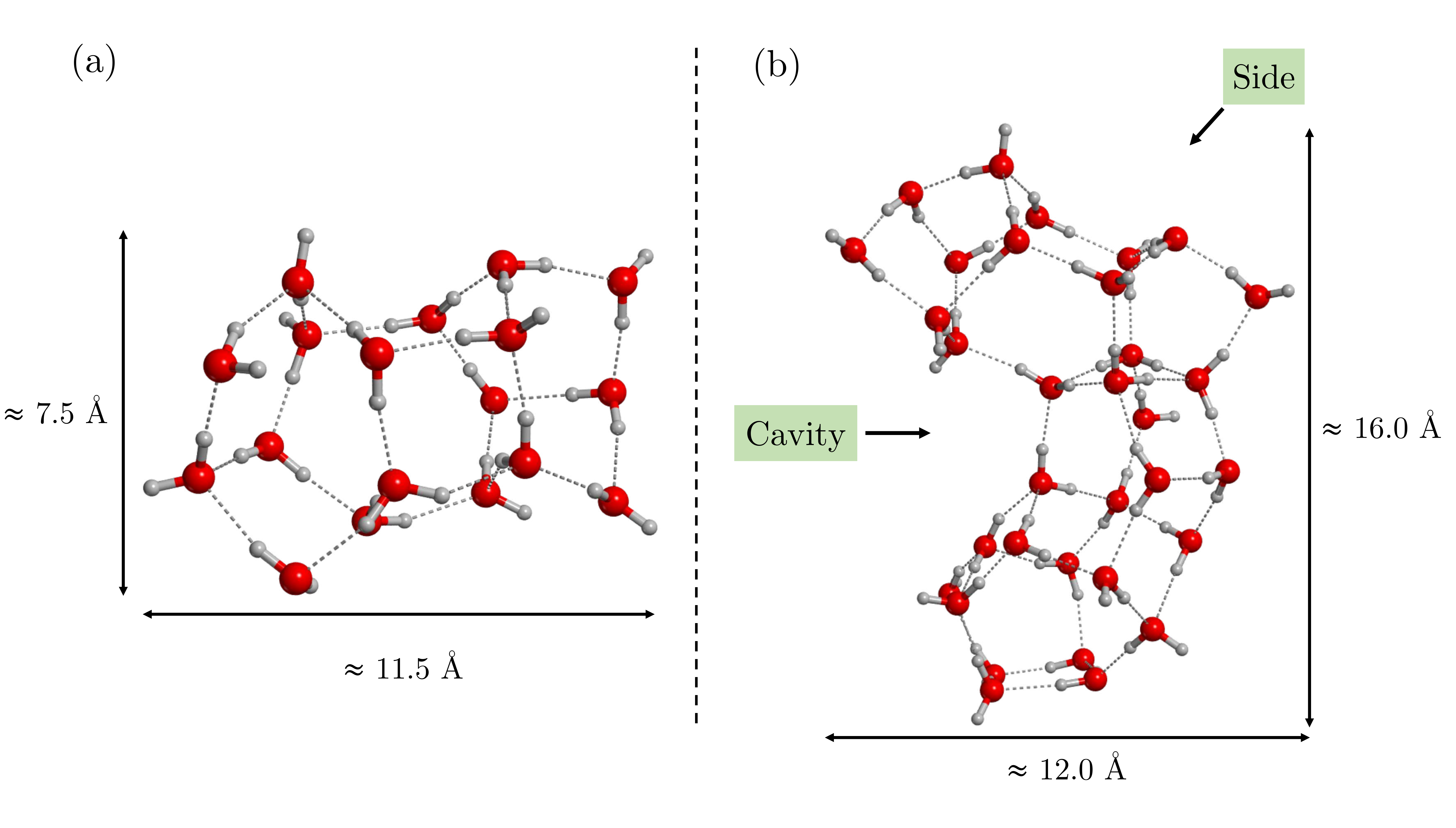}
    \caption{Structures of the 18 and 33 water molecules clusters, (a)
      and (b) respectively, optimised at the BHLYP/6-31+G(d,p) level.}
    \label{fig:w18-33}
\end{figure}

Radical-radical reactivity was also studied more realistically on two
amorphous solid water (ASWs) ices modelled by
molecular clusters consisting of 18 (W18) and 33 (W33) water
molecules, which were also used in previous
works.\cite{rimola2014,enrique2016,rimola2018}
The optimized structures are shown in Figure
\ref{fig:w18-33}. Interestingly, W33 exhibits a hemispherical cavity
and, accordingly, we studied the surface processes considering both
this cavity and an extended side of the ice surface (see Figure
\ref{fig:w18-33}(b)), as they exhibit different surface properties. In
order to make the calculations computationally affordable, for these
cases the DFT methods were combined with the Pople's 6-31+G(d,p) basis set. 

All stationary points were characterized by the analytical calculation
of the harmonic frequencies as minima (reactants, products and
intermediates) and saddle points (transition states). Intrinsic
reaction coordinate (IRC) calculations at the same level of theory
were carried out when needed to ensure that the transition states connect with the
corresponding minima.  Thermochemical
corrections to the potential energy values were carried out using the
standard rigid rotor/harmonic oscillator formulas\cite{mcquarrie2000}
to compute the zero point energy (ZPE) corrections.

Adsorption energies of HCO, CH$_3$ and NH$_2$ on W18 and W33 were
refined with single point energy calculations combining the DFT
methods with the extended 6-311++G(2df,2pd) basis set and corrected
for the basis set superposition error (BSSE). Considering \textit{A}
as the adsorbate and \textit{B} as the surface cluster model, the
BSSE-non-corrected adsorption energy was computed as
$\Delta E_{ads} = E^{AB}_{AB}(AB) - E^{A}_A(A)- E^{B}_B(B)$, where
superscripts denote the basis set used and the subscripts the geometry
at which the calculation was done. BSSE-corrected energies were
calculated as:

\begin{equation}
    \Delta E_{ads}^{CP} (AB) = \Delta E_{ads} + {BSSE}(A) + {BSSE}(B) + \delta^A(A) + \delta^B(B)
    \label{eq:binding_bsse}
\end{equation}

where the BSSE values were calculated following the Boys and Bernardi counterpoise correction method ($BSSE(A)=E^{AB}_{AB}(A)-E^{A}_{AB}(A)$, \cite{boys1970}), and where the deformation of each monomer was also  accounted for ($\delta^A(A)=E^A_{AB}(A)-E^{A}_A(A)$). 

Inclusion of ZPE corrections allowed us to obtain adsorption enthalpies at 0 K: 
\begin{equation}
    \Delta H_{ads}(AB) = \Delta E_{ads}^{CP} (AB) + \Delta ZPE
    \label{eq:bind_bsse_enth}
\end{equation}
In the sign convention followed in this work, the adsorption energy is the negative of the binding energy, i.e. $\Delta E_{ads} = -\Delta E_{bind}$. \\

The systems containing two radical species were first optimized in the
triplet electronic state, which was then optimized in the singlet
state to describe the biradical system. Structures with doublet and
triplet electronic states were simulated with open-shell calculations
based on the unrestricted formalism. Singlet biradicals systems were
calculated adopting the unrestricted broken symmetry (BS) approach, in
which the most stable initial wave function was found using the
\textit{stable=opt} keyword in Gaussian09.

Following the International System Units, all energy units are given
in kJ mol$^{-1}$, whose conversion factor to K is 1 kJ mol$^{-1}=$
120.274 K.

\section{3. Results}

In this section, results of the calibration study devoted to the
reactivity and interaction of HCO with CH$_3$ and NH$_2$ in the
presence of 1 and 2 water molecules are first presented. Then, results
on adsorption properties and the radical-radical reactions on W18 and 
W33 are reported.

\subsection{3.1. Methodology benchmark}
This section aims to be a calibration study to check the reliability
of the B3LYP and BHLYP methods for (i) the radical-water interactions 
and (ii) the activation energy of the reactions of CH$_3$/HCO and 
NH$_2$/HCO, in both cases with one and two water molecules (W1 and 
W2 hereafter) as minimal models representing an ice surface.
The references used for the study of the binding and activation energies
are CCSD(T) and CASPT2, respectively. In all cases, single point energies were
computed onto the B3LYP-D3 optimised geometries. In the supporting 
information (SI) section the reader can find these geometries as well as
the relative errors of the data presented in the following.

Table \ref{tab:InterEner} contains the binding
energies for the three radicals involved 
in this study (namely CH$_3$, NH$_2$ and HCO) interacting with W1 and W2.
The systems where a radical interacts with W1 have been based  on those
presented by \citet{wakelam2017}, where the initial structures were built in a
chemical-wise manner following the ability of each component of the radical
to establish a hydrogen bond to a single water molecule, e.g. in
the cases where NH$_2$ and HCO interact with W1 two possibilities 
were considered: the radicals acting as either H-bond donors 
(through one of their H atoms), or acting as H-bond acceptors 
(through the N and O atoms, respectively).
The initial geometries of the systems with W2 were built similarly,
with the radicals having the maximum number of hydrogen bonds to the
two water molecules.\\

\begin{table}
\centering
\caption{ZPE- and BSSE-non-corrected binding energies computed with 
different methods for the interaction of CH$_3$, HCO and NH$_2$ with 1 and 2
water molecules (W1 and W2) at the B3LYP-D3 and BHLYP levels. None, D2 (accounting for 2-body 
interactions) and D3 (accounting for 2- and 3-body interactions) dispersion 
corrections were considered for the latter.
BHLYP-based and CCSD(T) values were calculated as single points on the B3LYP-D3 optimised
geometries.
Energy units are in kJ/mol.}
\label{tab:InterEner}
\begin{tabular}{|c|c|c|c|c|c|} 
\cline{2-6}
\multicolumn{1}{c|}{} & B3LYP-D3    & BHLYP        & BHLYP-D2    & BHLYP-D3    & CCSD(T)  \\ 
\hline
CH$_3$/W1             & 9.5      & 5.2   & 11.3     & 8.1      & 6.9      \\ 
\hline
HCO/W1 (H)            & 12.9     & 10.9  & 13.6     & 13.1     & 12.0     \\ 
\hline
HCO/W1 (O)            & 11.7     & 9.7   & 12.5     & 12.1     & 11.5     \\ 
\hline
NH$_2$/W1 (H)         & 14.1     & 12.0  & 14.9     & 14.4     & 12.8     \\ 
\hline
NH$_2$/W1 (N)         & 23.4     & 21.5  & 24.3     & 23.6~    & 21.6     \\ 
\hline
CH$_3$/W2             & 8.6      & 3.1   & 11.3     & 8.0      & 7.1      \\ 
\hline
HCO/W2                & 14.8     & 9.2   & 16.1     & 15.1     & 13.2     \\ 
\hline
NH$_2$/W2             & 41.6     & 36.6  & 44.5     & 41.9     & 38.3     \\
\hline
\end{tabular}
\end{table}

It can be seen that
    CH$_3$ is the species having the weakest interaction with the
    water molecules (6.9-7.1 kJ/mol at the CCSD(T) level). HCO and NH$_2$,
    instead, can form intermediate and strong H-bonds respectively,
    and, accordingly, they show higher binding energies (11.5-13.2
    kJ/mol for the former and 12.8-38.3 kJ/mol for the latter at the
    CCSD(T) level).  This trend is in agreement with that found by
    \citet{wakelam2017}, in which the binding energies of these species
    with W1 were 13, 19-23 and 23-38 kJ/mol for CH$_3$, HCO and NH$_2$ respectively, computed at the M062X/aug-cc-PVTZ level. Note, however, that the quoted values are not the final values used in the model by \citet{wakelam2017}, as they also use binding energies from other sources in the literature in order to fit experimental data.
    \\
    Regarding the performance of the methods, the best ones are
    B3LYP-D3 and BHLYP-D3, with a relative error of 2-39\% and 5-18\% 
     respectively (see SI section).
     The pure BHLYP method systematically 
     underestimates the binding energies, providing strong deviations for 
     CH$_3$ and HCO, specially on W2. BHLYP-D2 dramatically overestimates the
     binding energies of CH$_3$-containing systems (relative errors of $\sim$ 60\%), indicating
     that the D2 term probably accounts for dispersion in excess for this
     kind of weakly bound complexes. 
     \\

Table \ref{tab:calibration} shows the calculated energy barriers
     of CH$_3$/HCO and NH$_2$/HCO on W1 and W2.
     Three different possible reactions have been identified: i)
     radical-radical coupling, leading to formation of the iCOMs
     (i.e. CH$_3$CHO and NH$_2$CHO); ii) direct hydrogen abstraction, in
     which the H atom of HCO is transferred to the other radicals, forming
     CO + CH$_4$ and CO + NH$_3$, respectively; and iii) water assisted hydrogen
     transfer, the same as ii) but the H transfer is assisted by the water
     molecules adopting a H relay mechanism. These reactions will be
     referred to as \textit{Rc}, \textit{dHa} and \textit{wHt}, respectively, along the
     work.

\begin{table}
\centering
\caption{ZPE-non-corrected energy barriers ($\Delta E^{\ddagger}$),
  computed with different methods, for radical-radical coupling (Rc),
  direct hydrogen abstraction (dHa) and water assisted hydrogen
  transfer (wHt) reactions of NH$_2$ + HCO and CH$_3$ + HCO in the
  presence of 1 and 2 water molecules computed at the B3LYP-D3 and BHLYP-based levels. 
  For the latter two dispersion corrections have been considered: D2 
  (which considers all 2-body interactions) and D3 (which considers 2- and 3-body interactions).
  BHLYP-based, CCSD(T) and CASPT2 values were calculated as single point energy 
  calculations on the B3LYP-D3 optimised geometries. 
  Energy units are in kJ mol$^{-1}$. NB stands for ``No
  Barrier" and means that the process is found to be barrierless.}
\label{tab:calibration}
\begin{tabular}{|c c|c|c|c|c|c|c|}
\hline
System \rule{0pt}{1cm}& Process & \shortstack{\textit{B3LYP}\\-\textit{D3}} &  \textit{BHLYP} & \shortstack{\textit{BHLYP}\\-\textit{D2}} & \shortstack{\textit{BHLYP}\\-\textit{D3}} & \textit{CCSD(T)} & \textit{CASPT2} \\ \hline
\multirow{3}{*}{\shortstack{NH$_2$/HCO\\$\cdots$ W1}} & 
Rc & NB & NB & NB & NB & NB & NB \\
 & dHa & 3.5  & 6.1 & 4.1 & 5.1 & 5.3 & 4.2 \\
 & wHt & 10.9 & 48.6 & 46.2 & 48.8 & 43.8 & 48.3 \\ \hline
\multirow{3}{*}{\shortstack{CH$_3$/HCO\\$\cdots$ W1}} & 
Rc  & NB & NB & NB & NB & NB & NB \\
 & dHa & 3.1 & 6.9 & 3.3 & 5.1 & 5.0 & 1.5 \\
 & wHt & 23.5 & 59.4 & 55.5 & 58.6 & 60.0 & 52.2 \\ \hline
\multirow{3}{*}{\shortstack{NH$_2$/HCO\\$\cdots$ W2}} & 
Rc & 6.5 & 8.5 & 6.1 & 6.8 & 6.3 & 6.1 \\
 & dHa & NB & NB & NB & NB & NB & NB \\
 & wHt & 15.5 & 65.8 & 62.5 & 65.1 & 52.3 & 52.9 \\ \hline
\multirow{3}{*}{\shortstack{CH$_3$/HCO\\$\cdots$ W2}} & 
Rc & NB & NB & NB & NB & NB & NB \\
 & dHa & 1.6 & 4.0 & 1.5 & 1.6 & 1.6 & 1.0 \\
 & wHt & 30.7 & 65.8 & 62.5 & 65.1 & 64.0 & 52.0 \\ \hline
\end{tabular}
\end{table}

According to these values, two general
    trends are observed: (i) Rc and dHa are either barrierless
    (meaning that the initial biradical structures were not stable) or
    have relatively low energy barriers (1.0-6.1 kJ/mol for
    CASPT2), and (ii) wHt are the processes presenting the highest 
    energy barriers (around 50 kJ/mol for CASPT2).

For those Rc and dHa reactions having an energy barrier, one can
     see that the worst performance is given by the pure BHLYP method compared to CASPT2. The 
     rest of the methods perform similarly, BHLYP-D2 and B3LYP-D3 being the best ones, which are  
     followed by BHLYP-D3.
     Regarding the wHt reactions, B3LYP-D3 dramatically underestimates the
     energy barriers (errors of $\sim$ 40-80\%), while BHLYP-based methods perform reasonably well with errors below $\sim$ 25\%.
     CCSD(T) performs similarly to
     the BHLYP-based methods,  presenting moderate energy barrier
     deviations compared to the CASPT2 values. All errors can be found in the SI section.

In summary, BHLYP-D3 provides the
    most reliable binding energies, B3LYP-D3, BHLYP-D2 and
    BHLYP-D3 show good performances for energy barriers related to
    Rc and dHa processes, while BHLYP-D2 and BHLYP-D3 perform good
    for wHt energy barriers. Accordingly, and with the aim to be consistent, in the following sections (devoted to
    the binding and potential energy profiles on the W18 and W33 cluster
    models) we provide the results at the BHLYP-D3 level of theory, while results based on BHLYP-D2 (on both W18 and W33) and on B3LYP-D3 (on W18) are reported in the SI.

\subsection{3.2. Radical-surface binding enthalpies}

Here we present the ZPE- and BSSE-corrected binding energies of the radicals on the
\textit{W18} and \textit{W33} cluster models and their comparison with
recent literature. These values are important because binding energies
are essential parameters in astrochemical modelling studies.

\begin{figure}[H]
    \centering
    \includegraphics[width=\textwidth]{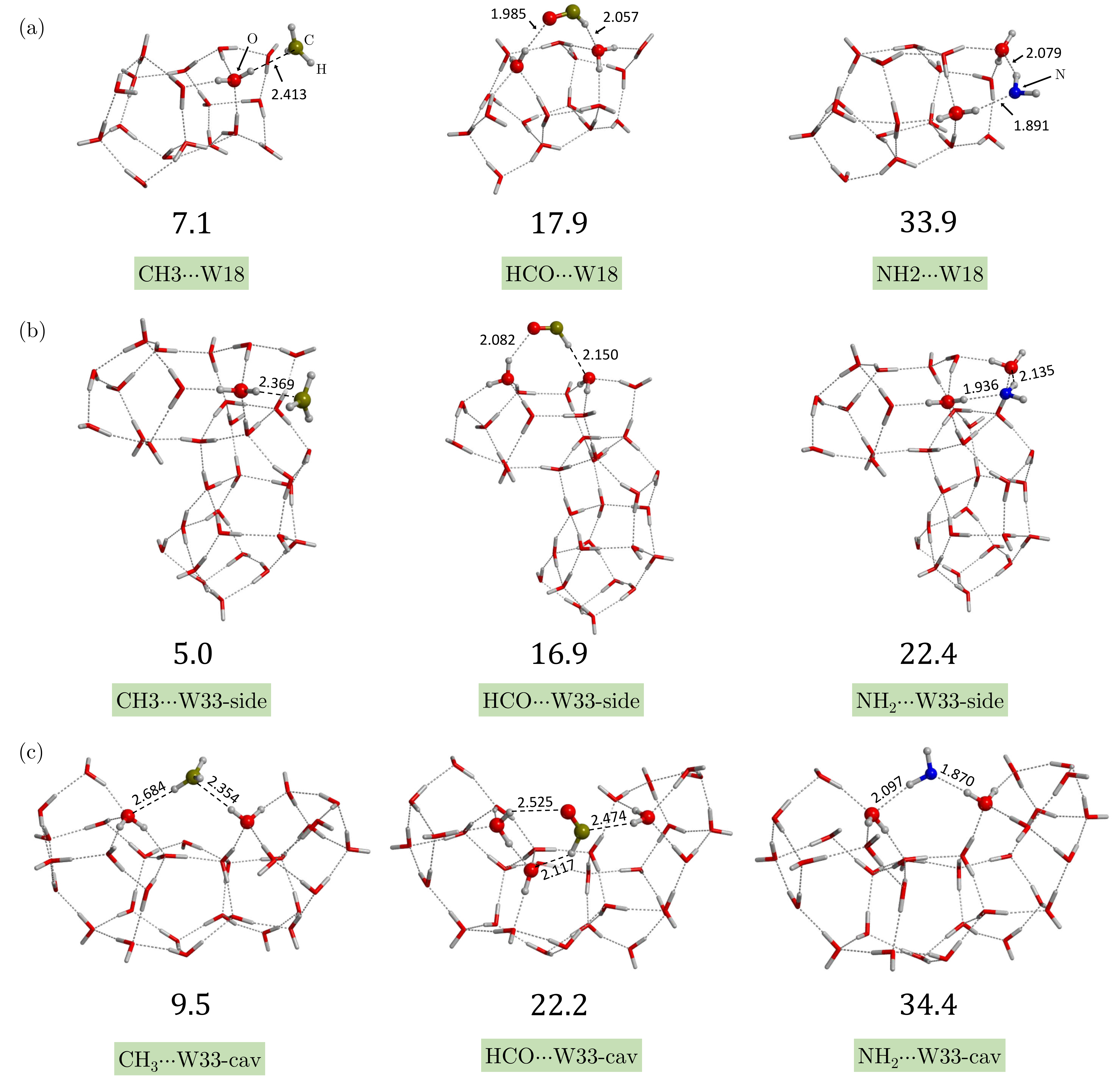}
    \caption{Fully optimized geometries for the binding of CH$_3$, HCO and
      NH$_2$ on (a) W18, (b) W33-side and (c) W33-cav at BHLYP-D3/6-31+G(d,p).
      Bond distances are in \r{A}.
      Binding enthalpies (at 0 K) values (in kJ/mol) are corrected for BSSE
      and are shown below their respective structures.}
    \label{fig:adsorbed}
\end{figure}

As it was stated in the previous section, the binding of the radicals
on the water ice surfaces are mainly dictated by hydrogen bonds
(H-bonds) and dispersion interactions. The clusters exhibit several
potential binding sites (Figure \ref{fig:w18-33}) and, accordingly, 
different radical/surface complexes can exist. 
Here, for the sake of simplicity, we choose those
complexes in which the inter-molecular interactions between the two
partners are maximized.
The underlying
assumption is that radicals on the  water ice surfaces have had enough time to thermalize and establish the largest number of
inter-molecular interactions with the surface, as very probably is the
case in the interstellar conditions.
Figure \ref{fig:adsorbed} shows the BHLYP-D3 fully
optimised complexes on W18 and on the two differentiated W33 sites: its side (W33-side) and its cavity (W33-cav). The corresponding BSSE-corrected binding 
enthalpies (at 0 K) are also shown. The same information at BHLYP-D2 and B3LYP-D3 is available in SI.

The binding enthalpy
    trend is NH$_2$ $>$ HCO $>$ CH$_3$, irrespective of the cluster
    model and surface morphology, in agreement with the results of the
    previous section.
    Specifically, the binding enthalpy ranges are: 22.4-34.4, 
    16.9--22.2 and 5.0--9.5 kJ/mol for NH$_2$, HCO and CH$_3$, 
    respectively. These values compare well with those previously
    computed by \citet{enrique2016} for HCO
    and CH$_3$ on the W18 cluster (19 and 6 kJ/mol) and by 
    \citet{rimola2018} for NH$_2$ and HCO on the W33-cav cluster
    (33.5 and 17.5 kJ/mol).
    In the model presented in \citet{wakelam2017}, the authors reported
    binding energies of 27, 20 and 13 kJ/mol for NH$_2$, HCO and CH$_3$,
    respectively, in rough agreement with our results.
    \citet{Sameera2017} have recently computed the binding energies of
    HCO and CH$_3$ adsorbed on hexagonal ice slabs\footnote{Authors used either
      full DFT with a double-$\zeta$ basis set or the ONIOM approach
      combining DFT with force fields. They reported several values
      owing to the different adsorption modes of these two species
      adopting the two computation approaches. This allowed them to
      report the wide range of binding energies reported here.}.
    They found HCO binding
    energies ranging between 12--40 kJ/mol, and 11--25
    kJ/mol for HCO and CH$_3$, respectively. These values are similar
    or moderately larger than the values found in our W33-cav cluster.\\

Finally, it is worth mentioning that these complexes do not present
hemi-covalent bonds, as it was the case for CN in
\citet{rimola2018}. Attempts to identify this type of binding in
the current systems have been made but the initial structures
collapsed to the complexes presented here upon geometry
optimization. Therefore, for these systems, the interaction of the
radicals with the ice surfaces is essentially dictated by H-bonds and
dispersion forces.

\subsection{3.3. Radical-radical reactivity}

\begin{figure}[H]
    \centering
    \includegraphics[width=0.75\textwidth]{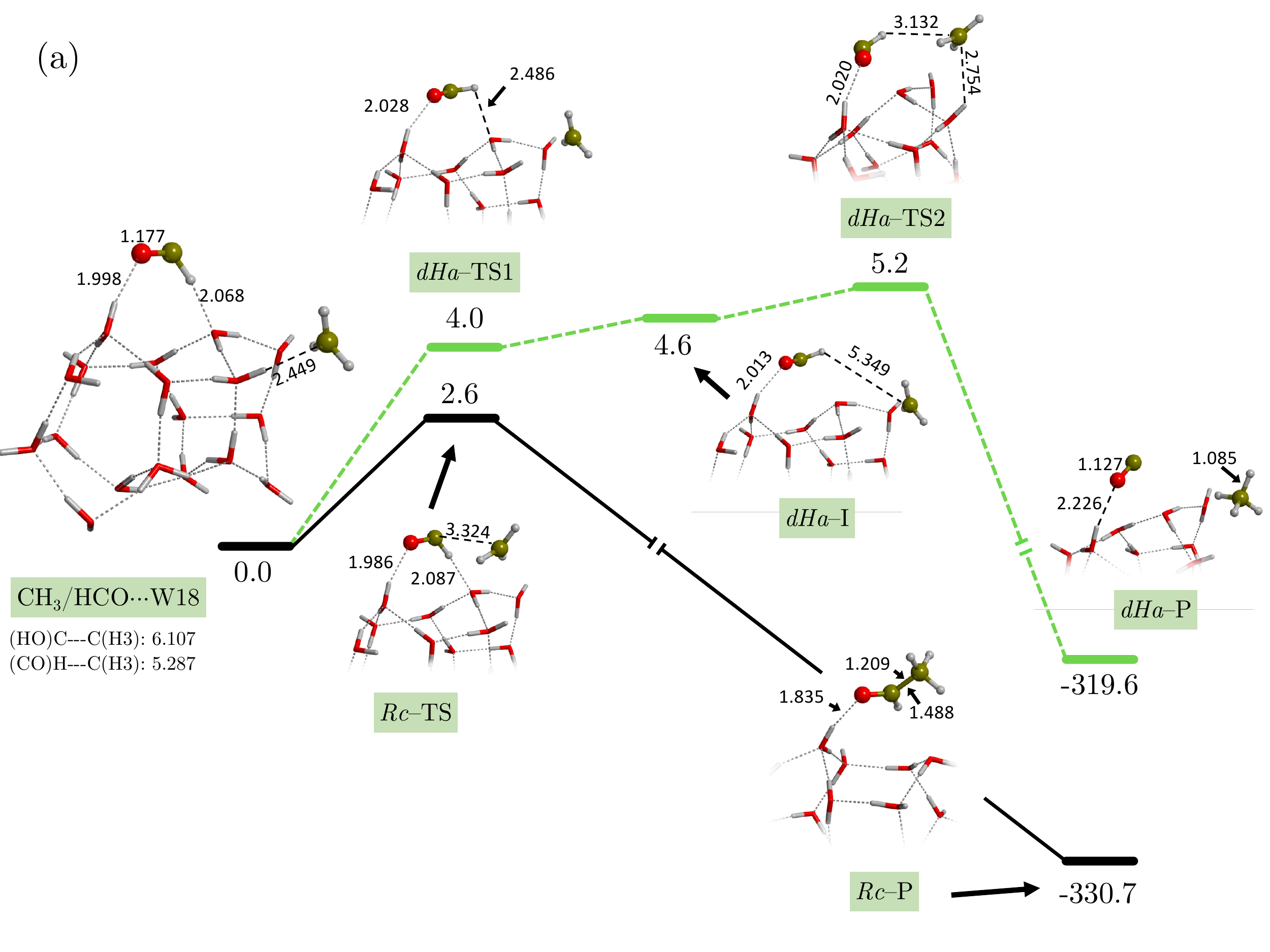}
    \includegraphics[width=0.75\textwidth]{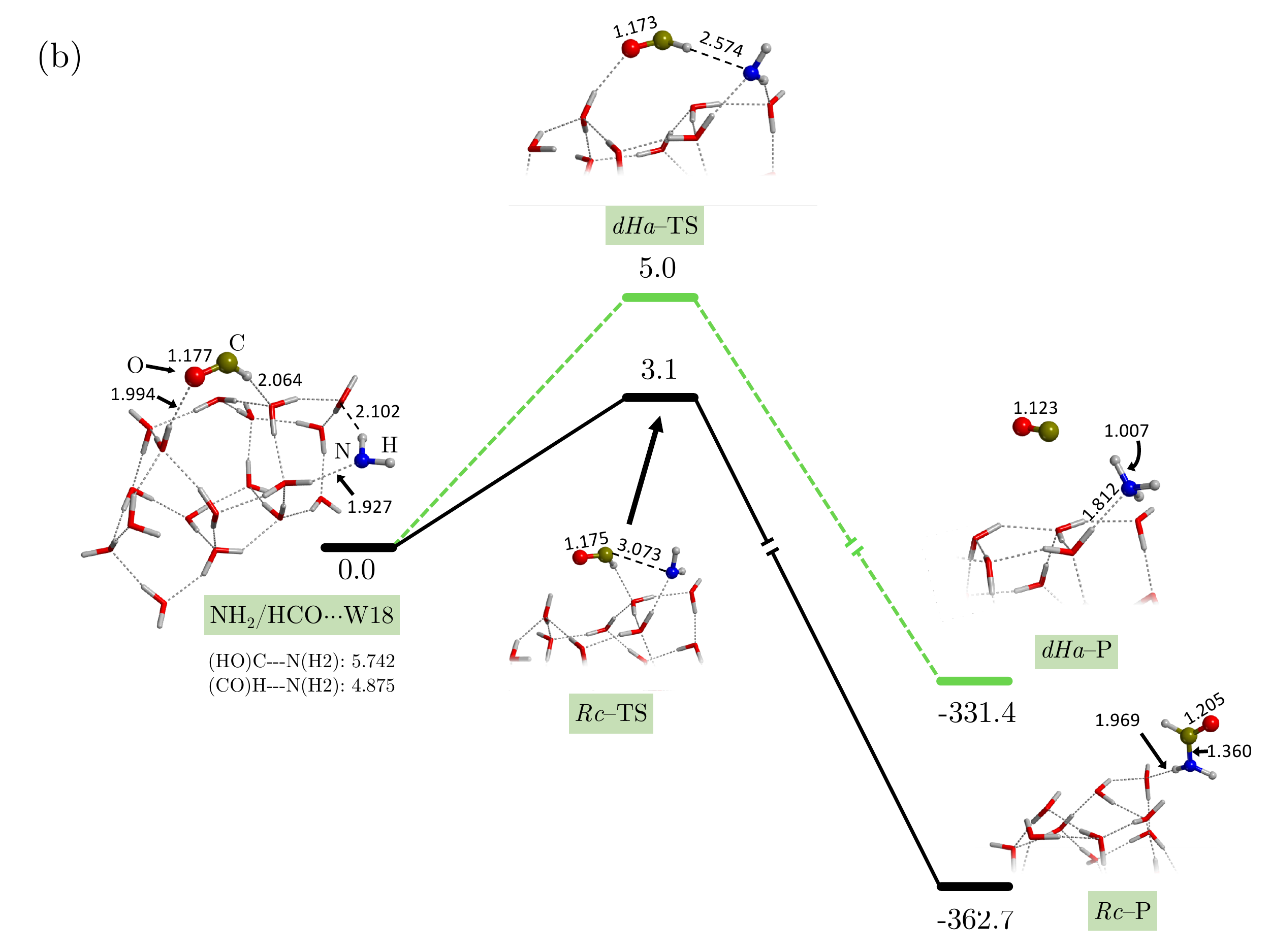}
    \caption{ZPE-corrected Rc (solid black line) and dHa (dashed green
    lines) PESs for (a) HCO/CH$_3$\,$\cdots$W18 and (b) 
    HCO/NH$_2$\,$\cdots$W18 calculated at the BHLYP-D3/6-31+G(d,p).
    Energy units are in kJ/mol and distances in \r{A}.
    Notice that dHa-TS1 for (a) HCO/CH$_3$\,$\cdots$W18 lies below the
    intermediate dHa-I, due to the ZPE correction, setting effectively
    a single energy barrier: dHa-Ts2.}
    \label{fig:w18_bhlyp}
\end{figure}

\begin{figure}[H]
    \centering
    \includegraphics[width=0.8\textwidth]{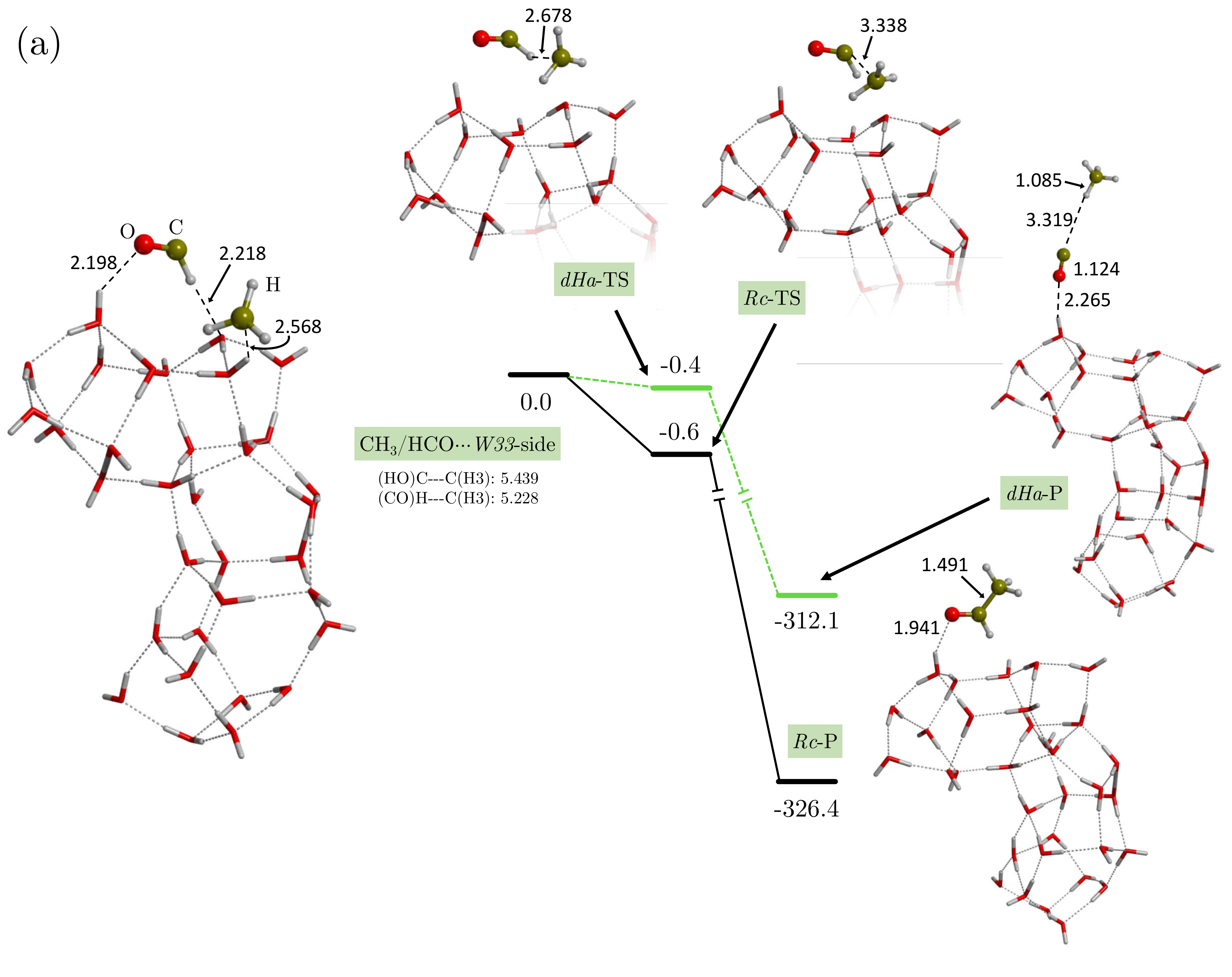}
    \includegraphics[width=0.8\textwidth]{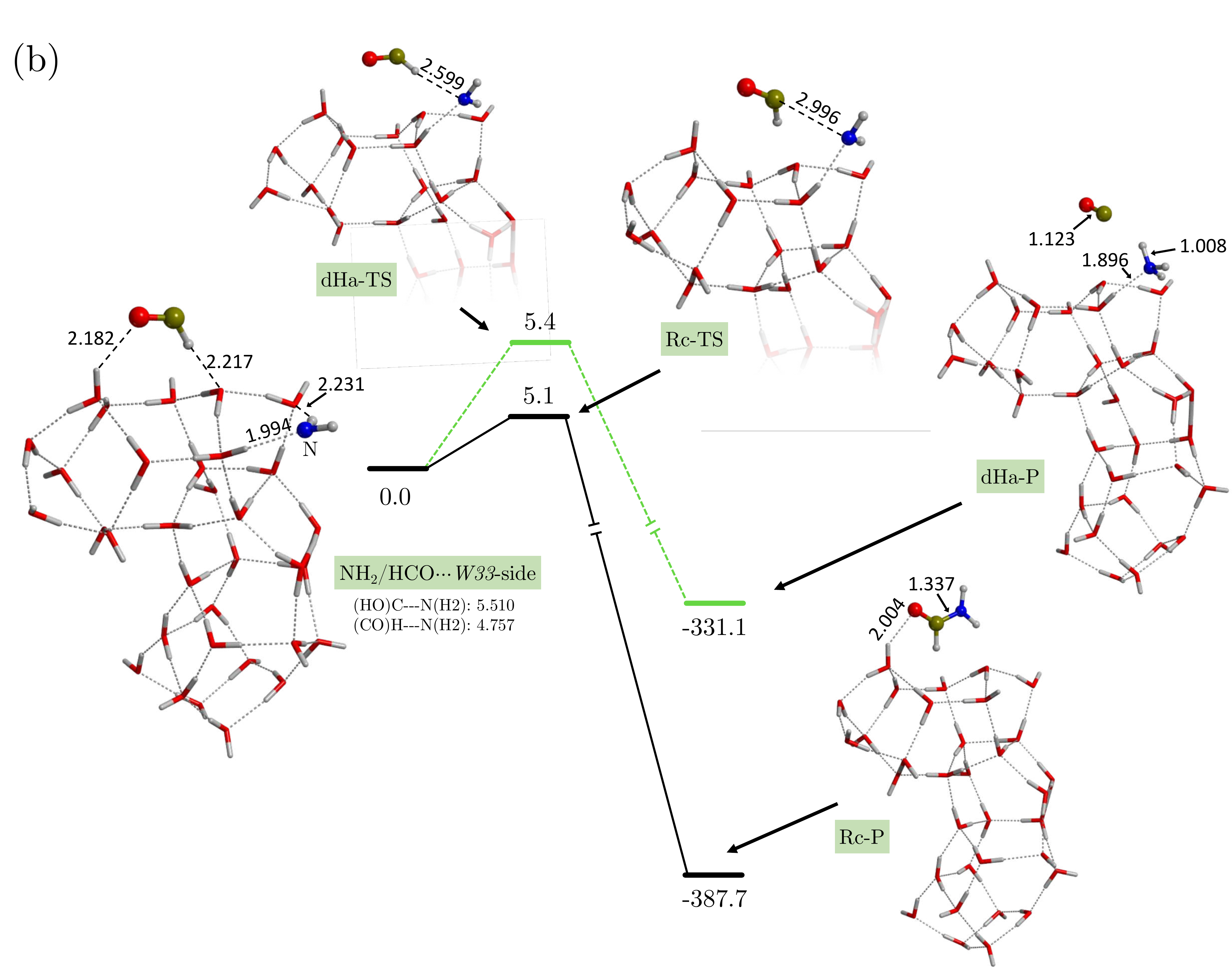}
    \caption{ZPE-corrected Rc (solid black lines) and dHA 
    (green dashed lines) PES for the HCO/CH$_3 \cdots$W33-side (a) 
    and HCO/NH$_2 \cdots$W33-side (b) systems optimized at the BHLYP-D3
    theory level. Energy units
    are in kJ/mol and distances in \r{A}.
    Notice that both Rc and dHa TS for CH$_3$/HCO lie below the
    energy of reactants due to the ZPE correction.}
    \label{fig:w33_Rc_dHa_side}
\end{figure}

\begin{figure}[H]
    \centering
    \includegraphics[width=0.84\textwidth]{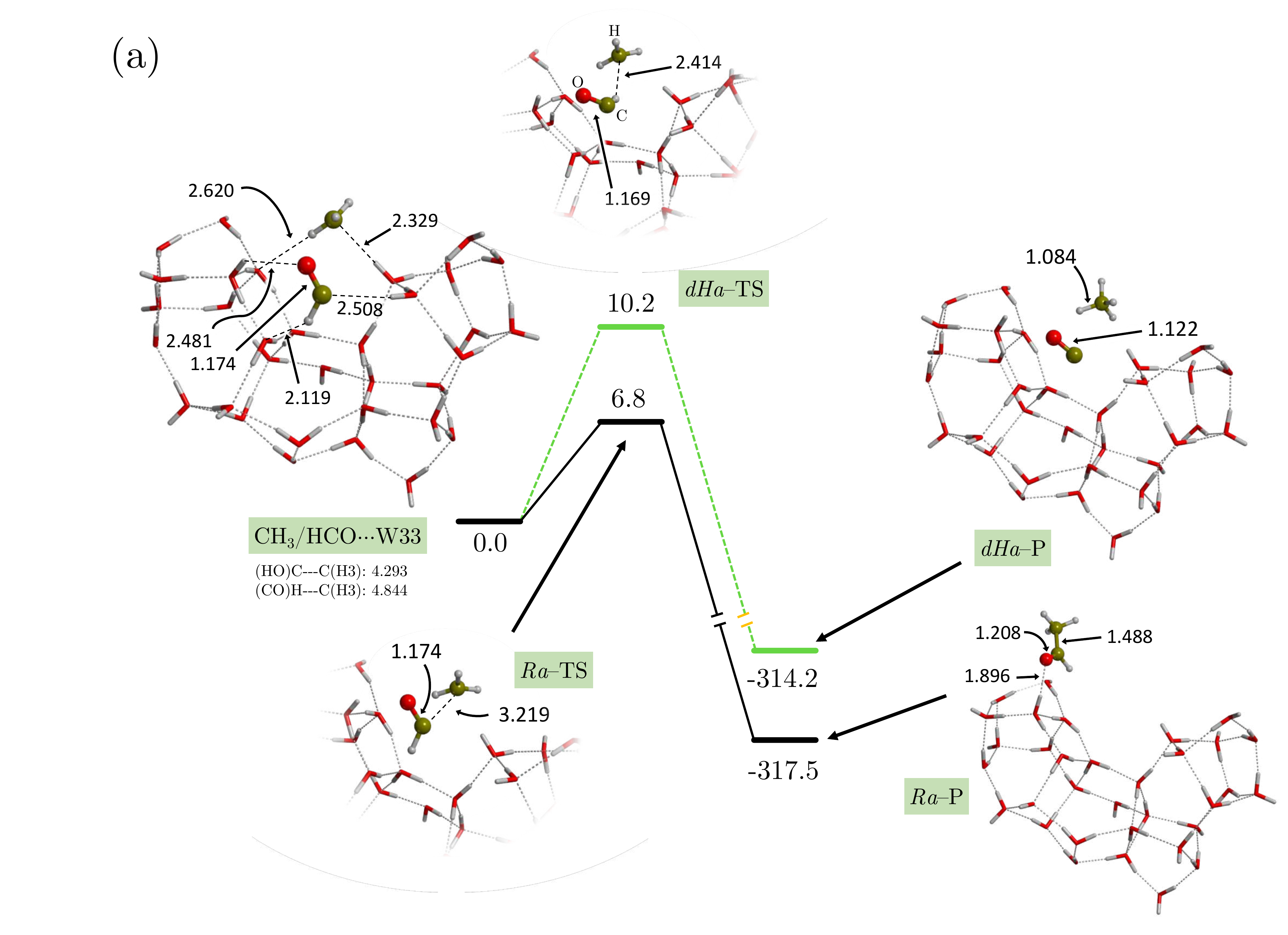}
    \includegraphics[width=0.84\textwidth]{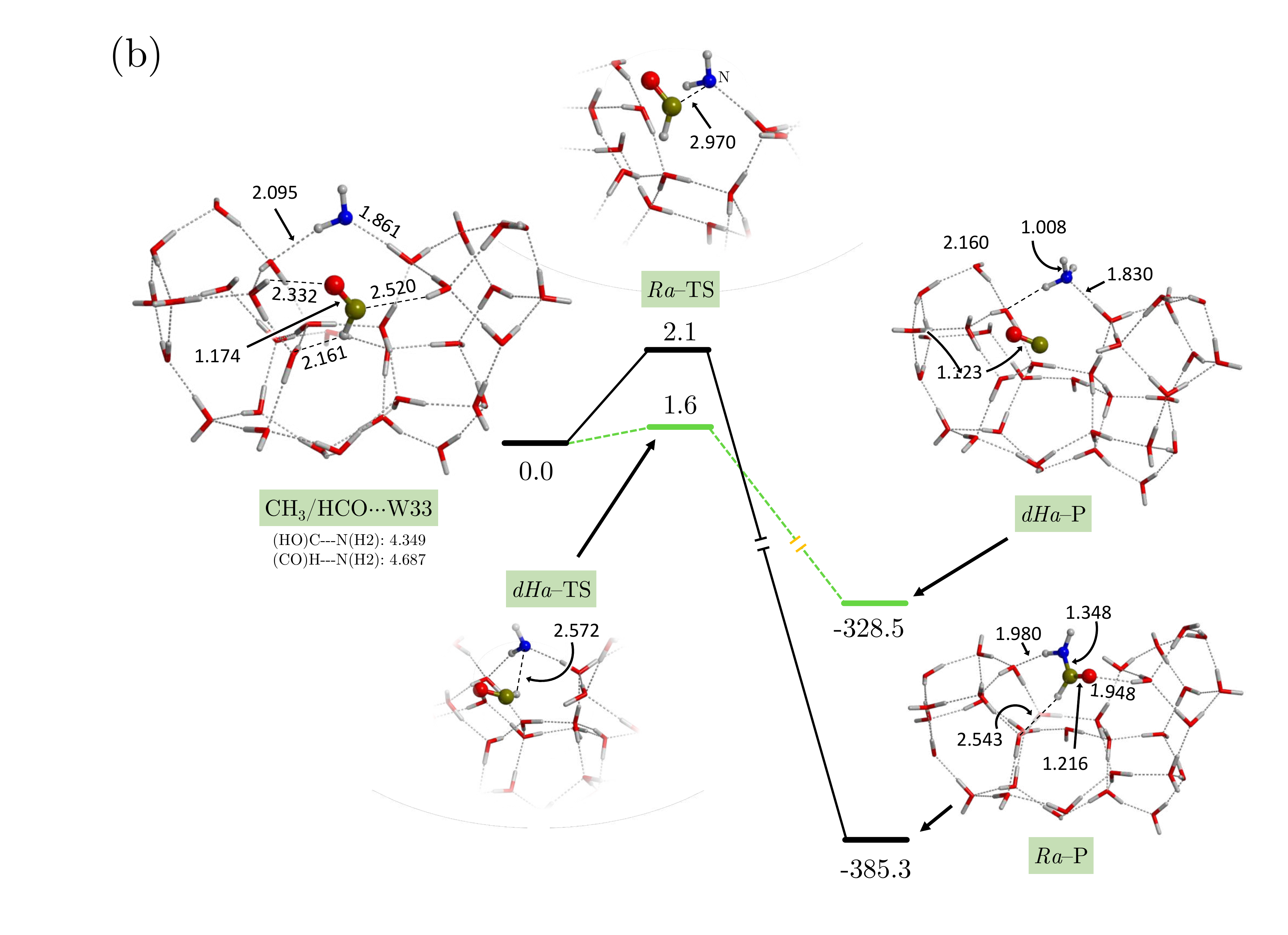}
    \caption{ZPE-corrected Rc (solid black lines) and dHA 
    (green dashed lines) PES for the 
    HCO/CH$_3\cdots$W33-cav (a) and  HCO/NH$_2\cdots$W33-cav (b) 
    systems optimized at the BHLYP-D3 theory level. 
    Energy units are in kJ/mol and distances in \AA.
    BHLYP-D2 values can be found in the SI.}
    \label{fig:w33_Rc_dHa_cav}
\end{figure}

In this section, the reactivity of HCO with CH$_3$ and NH$_2$ on W18, W33-side 
and W33-cav are investigated. Given that the chemical environment between
W18 and W33-side is similar (note that the surface morphology of W33-side
is very similar to that of W18 because W33 was built up by joining two W18
units\cite{rimola2018}), the comparison of the results between these two models will 
allow us to assess the effects introduced by the ASW model size.
On the other hand, comparison between the reactivity on W33-side and W33-cav will allow us to
assess the effects due to different surface environments, 
namely the presence of a higher number of radical/surface interactions. Please note that the
cavity sites are probably more representative of the interstellar
conditions than the side sites, as the vast majority of radicals are trapped in the bulk of the ice.\\
Figures \ref{fig:w18_bhlyp}, \ref{fig:w33_Rc_dHa_side} and \ref{fig:w33_Rc_dHa_cav} 
show the PESs of the reactivity of CH$_3$/HCO and NH$_2$/HCO calculated at BHLYP-D3 on top of W18, W33-side and W33-cav, respectively. Based on the results for the reactivity in the presence of W1 and W2 (see \S 3.1), 
the same three reaction paths, i.e., \textit{Rc}, \textit{dHa} and \textit{wHt}, have been investigated. However, as the later processes are those exhibiting the highest activation enthalpies (as high as 100.6, 78.8, 92.5 and 79.1 kJ/mol for CH$_3$/HCO$\cdots$W18, 
CH$_3$/HCO$\cdots$W33-cav, NH$_2$/HCO$\cdots$W18 and NH$_2$/HCO$\cdots$W33-cav, respectively), for the sake of clarity and with the aim of focusing only
on the reactions that might play a role in interstellar chemistry, all results related to wHt can be found in SI, this section only showing the
Rc and dHa reactions calculated at BHLYP-D3. In the same way, the reader can also find in SI the results for all
systems computed at BHLYP-D2 and, for W18, also at
B3LYP-D3.

The initial structures of these systems were built according to the interaction 
patterns present in the single adsorbed radical complexes (see reactant structures of Figures 
\ref{fig:w18_bhlyp}-\ref{fig:w33_Rc_dHa_cav}).
The Rc and dHa reactions, forming acetaldehyde or formamide
and CO+CH$_4$ or CO+NH$_3$, respectively, both take place
through a single step. That is, the bond formation
between the two radicals for Rc, and the H transfer for dHa.
The only exception is the dHa reaction of CH$_3$/HCO$\cdots$W18, which 
displays first a submerged activation energy step where HCO breaks its H-bonds with the surface to facilitate the H transfer (see dHa-TS1 of Figure \ref{fig:w18_bhlyp}(a)).
By comparing the Rc and dHa reactions on these three cluster models, the general trend is that Rc activation enthalpies are lower than the dHa ones, between 0.5 - 4 kJ/mol lower. This was already observed in the presence of W1 and W2, where most of the Rc reactions are barrierless while the dHa present physical (although low) energy barriers. 
It should be noticed that the HCO/CH$_3$ reactions on W33-side (Figure \ref{fig:w33_Rc_dHa_side}a) are barrierless due to the ZPE correction, as opposed to the the W1 and W2 cases where the lack of barrier is because of the unstable singlet evolving directly towards products. 
Indeed, all the stationary points shown in Figure \ref{fig:w33_Rc_dHa_side}(a) have been identified as
stable structures in the pure PESs (i.e.,
without ZPE corrections) but after introduction of ZPE
corrections, the transition state becomes lower in energy than
the reactants, and hence the barrierless character.

Another interesting point is that results for W18 and W33-side present some differences, particularly for the HCO + CH$_3$ reactions: on W33-side, both Rc and dHa are barrierless while on W18 they present energy barriers. This is indicative of the fact that the size of the cluster for this radical-radical reactivity is important. Finally, no clear trends related to the effect of the ASW ice morphology can be obtained by comparing the W33-side results with the W33-cav. That is, for HCO+CH$_3$ reactions, activation enthalpies are higher on W33, while the opposite occurs for HCO+NH$_2$ ones.\\
In the following section, a comprehensive discussion of all these results is provided.


\section{4. Discussion}

\subsection{4.1 Reaction channels and competition to iCOMs formation}

Our computations show that if HCO + CH$_3$ and HCO
+ NH$_2$ react on top of an ASW ice, they have two sets of
possible reaction products: i) formation of iCOMs (\textit{Rc}
process), where the two radical species meet and couple, and ii) the
formation of the hydrogenated CH$_4$/NH$_3$ species, where the H atom
of HCO migrates to CH$_3$/NH$_2$, which can happen either directly
(\textit{dHa} process) or through the ice water molecules adopting a H
transfer relay mechanism (\textit{wHt} process).  The energetics
associated with each process at the BHLYP-D3 level are summarised in Table
\ref{tab:summary_energies}. 

\begin{table}
\centering
\caption{Highest activation enthalpies (at 0 K) for \textit{Rc}, 
\textit{dHa} and \textit{wHt} reactions on W18, W33-side and W33-cav
at the BHLYP-D3 level. Note that NB stands for ``No Barrier''. Units are in kJ/mol.}
\label{tab:summary_energies}
\begin{tabular}{|c|ccc|ccc|ccc|} 
\cline{2-10}
\multicolumn{1}{c|}{}                                  & \multicolumn{3}{c|}{W18} & \multicolumn{3}{c|}{W33-side} & \multicolumn{3}{c|}{W33-cav}  \\ 
\hline
Sys                                                    & Rc  & dHa & wHt          & Rc  & dHa & wHt               & Rc  & dHa  & wHt              \\ 
\hline
HCO + CH$_3$ & 2.6 & 5.2 & 100.6        & NB  & NB  & -                 & 6.8 & 10.2 & 78.8             \\ 
\hline
HCO + NH$_2$  & 3.1 & 5.0 & 92.5         & 5.1 & 5.4 & -                 & 2.1 & 1.6  & 79.1             \\
\hline
\end{tabular}
\end{table}


The \textit{Rc} and \textit{dHa} reactions show, in all the studied
systems, similar energetic features, i.e., they are either barrierless
or exhibit relatively low energy barriers. The highest pair of 
energy barriers concerns the HCO + CH$_3$ reactions on W33-cav, i.e., 6.8 and 10.2
kJ/mol for \textit{Rc} and \textit{dHa}, respectively.
As a general trend, \textit{dHa} reactions have slightly higher activation
energies than \textit{Rc} (by as much as 3.4 kJ/mol for HCO/CH$_3$
and 1.9 kJ/mol for HCO/NH$_2$).
In some cases, like HCO/CH$_3\cdots$W33-side, HCO/NH$_2\cdots$W33-side and
HCO/NH$_2\cdots$W33-cav both Rc and dHa can be considered as competitive 
reactions given the small activation energy differences.
In contrast, the lowest energy barriers for wHt HCO + CH$_3$ and HCO + NH$_2$
reactions are $\sim$ 80 kJ/mol, respectively. These values are larger 
than any \textit{Rc} and \textit{dHa} energy barrier and, accordingly, 
\textit{wHt} reactions cannot be considered by any means as competitive
channels. The explanation of these energetic differences is provided by the reaction mechanisms. Rc and
dHa reactions take place, in most of the cases, in a
concerted way, in which the radicals, in essence, have to
partly break the interactions with the surface to proceed
with the reaction. In contrast, most of the \textit{wHt}
reactions adopt a multi-step mechanism since the H transfer, which is assisted by different ice water molecules, involves different breaking/formation bonds. In these cases, high energy intermediates consisting of the coexistence of HCO and an OH radical are involved (see SI).\\

In previous works by some of us, (e.g. \cite{enrique2016}) the
\textit{wHt} reactions between CH$_3$ + HCO on the W18 cluster model
were observed to spontaneously occur during geometry optimisation,
i.e. they were found to be barrierless.  The difference with the
computations presented in this work resides on the fact that, in the
previous work\cite{enrique2016}, the initial wave function did not
describe a singlet biradical system but a metastable singlet
closed-shell-like one, and hence the spontaneous evolution to form
CO+CH$_4$. In this work, as well as in
  Rimola et al. (2018)\,\cite{rimola2018}, the initial wave function
is actually describing a singlet biradical situation, which leads to a
significant stabilization of the reactants and hence the presence of
high energy barriers.\\

Finally, some words related to the chemical role played by the ice on
these reactions are here provided. The \textit{Rc} and \textit{dHa}
processes in the gas-phase (namely, in the absence of the icy grain) are,
in both cases, barrierless. In contrast as explained above, in the presence of the
surface, they exhibit, although low, energy barriers. Accordingly,
from a rigorous chemical kinetics standpoint, the grains slow down the
reactions. This leads us to think that a major role played by the
grains is as that of third bodies, by quickly absorbing the nascent energy
associated with the reactions, hence stabilising the products. This aspect is
particularly appealing in the iCOMs formation processes via radical recombination since in the
gas phase iCOMs can redissociate back to reactants if they are not
stabilised through three-body reactions. As we will discuss below, the
water morphology plays a role in the reaction energetics, although it
does not change the essence of our conclusions. 

\subsection{4.2 Influence of the water ice surface model}

Clear differences arise when comparing W33-side and W33-cav, as radicals on the latter exhibit more
inter-molecular interactions with the surface.
This can be seen for example in the binding energies (higher on W33-cav 
than on W33-side, see Figure \ref{fig:adsorbed}). But also on the Rc and dHa energy barriers, for which
a different behaviour is observed. Indeed, for HCO + CH$_3$ these reactions are barrierless on W33-side, while on W33-cav they present energy barriers of 7 (Rc) and 10 (dHa) kJ/mol.
On the other hand, for HCO + NH$_2$ the opposite behaviour is observed: the energy barriers are higher on W33-side than on W33-cav (see Table \ref{tab:summary_energies}).
This might be indicating that the different polarity of the radicals, i.e., CH$_3$/apolar and NH$_2$/polar, is important when several polar-based inter-molecular interactions surround the reaction sites, like it is the case of W33-cav.\\

Finally, the size difference between the W18 and W33 models does not seem to provide a consistent trend, neither for binding energies nor for the PESs.
This is probably due to the modest energetics of reactions of interest, which are the result of many intermingled effects, i.e. H-bond and dispersion interaction strength, small charge transfer and polarization. All these components are affected by the nature and size of the water adopted clusters without a definite and predictable structure-properties relationship.

\subsection{4.3 Astrophysical implications}

A major goal of this study is to understand whether iCOMs can be
formed on the icy grain surfaces by the direct combination of
radicals, a process assumed to be efficient in the majority of current
astrochemical models (see Introduction). The present computations show
that (i) there is a feasible channel leading to iCOM
  formation through radical-radical combination, (ii)
this channel may possess a barrier, and (iii) there
is at least a competitive reaction where radicals exchange a hydrogen
atom, the outcome of which is somewhat a step backwards in chemical
complexity as the products are simple hydrogenated species (CH$_4$ and
NH$_3$ in the current work) and CO.

The present computational data does not allow us to
definitively exclude the presence of a barrier in the radicals
combination. Indeed, although common sense would indicate a lack of barrier, calculations show that the
presence of the ice water molecules introduces an inter-molecular interaction
that depends on where the radicals are placed and on the radical polarity. 
This interaction probably necessitates energy to be broken: it is not obvious that
this energy is available in the ISM environments.
In fact, according to Garrod et al. model \cite{garrod2006}, once the
radicals are formed, they remain frozen on the ice and subsequently
more ice layers build up on top. The radicals remain imprisoned in
cavity structures inside the mantle and once the temperature reaches
$\sim$ 30 K due to the evolution of the central protostar, they
diffuse and react.  Among the 3 reactions sites discussed in this
article (W18, W33-side and W33-cav), the one best resembling this
picture is given by W33-cav (see Figure \ref{fig:w18-33}) due to the
larger number of inter-molecular interactions.  The Rc and dHa energy
barriers in the cavity have been shown to be larger for HCO + CH$_3$
than for HCO + NH$_2$, due probably to the different polarity of
CH$_3$ and NH$_2$.  If converted to Kelvin, the Rc and dHa barriers
for HCO + CH$_3$ and HCO + NH$_2$ on W33-cav are about 800, 1200,
  250 and 190 K, respectively (see Table \ref{tab:summary_energies}).
Thus, the efficiency of
these reactions is not expected to be very high, especially for HCO +
CH$_3$.

It has also to be noted that the starting points from where we study
the reactivity in W33-cav (see reactant geometries from Figure
\ref{fig:w33_Rc_dHa_cav}) contain both radicals very close by and in
the same cavity site (given the computational cost of higher quality
calculations we cannot simulate much larger clusters). However, in a
more realistic situation each radical would be stored in different
cavities and thus the actual barriers to overcome would not only
involve breaking the radical/ice inter-molecular interactions, but also
surmounting the ice surface diffusional barriers,
decreasing in this way the efficiency of Rc and dHa reactions, even
if they were ultimately barrierless.


We conclude this part mentioning that astronomical observations can
also bring useful constraints to the formation routes of iCOMs
  showing alternative routes to the ones explored in this study. For
  example, high spatial resolution observations of formamide line
emission towards the protostellar shock site L1157-B1 have
demonstrated that the formation of formamide is dominated by the
gas-phase reaction NH$_2$ + H$_2$CO \cite{Codella2018SOLIS} , a
reaction theoretically studied by some of us
\cite{Barone2015Formamide, Skouteris2017Formamide}.  On the same line,
observations of the deuterated forms of formamide (namely containing D
rather than H atoms) also provide strong constraints on the formation
route of this species in the hot corino of the solar-type protostar
IRAS 16293-2422.  The comparison of the measured NHDHCO/NH$_2$CHO and
NH$_2$CDO/NH$_2$CHO abundance ratios\cite{Coutens2016} with those
predicted by theoretical quantum chemical
calculations\cite{Skouteris2017Formamide} strongly favors a gas-phase
origin of formamide also in this source. Therefore, it is very
  likely that both grain-surface and gas-phase reactions contribute to
the enrichment of iCOMs in the ISM, playing different roles in
different environments.

\section{5. Conclusions and perspectives}

In this work, we have carried out an accurate study of the chemistry
of two couples of radicals, HCO + CH$_3$ and HCO + NH$_2$, on icy
surfaces. Our goal was to understand the possible reactions between
the two radicals on water ice mantles, and how the results depend on
the accuracy of the employed quantum chemical methods and on the
adopted surface models. To this end, we used different quantum
chemistry methods, in particular two hybrid DFT methods, B3LYP and
BHLYP, plus the wave function based CCSD(T) and multi
reference-based CASPT2 ones.  In addition, we
adopted different cluster models simulating the water surfaces: we
started with the simple cases of one and two water molecules to
identify the basic processes and to test the methodology, and then using
two different, large molecular cluster models for the ASW surfaces, of 18 and 33 water
molecules, respectively.

The conclusions of this work are the following:\\
1. If the reaction occurs, two channels are
possible: (i) the combination of radicals into acetaldehyde/formamide
and (ii) the formation of CH$_4$/NH$_3$ plus CO, where the H atom of
HCO is passed to CH$_3$/NH$_2$ via H abstraction.\\
2. The two reaction channels are either barrierless or
have relatively low energy barriers, from about 2 to 10 kJ/mol, as summarized in Table \ref{tab:summary_energies}.\\
3. Comparison of the results obtained with B3LYP-D3 and BHLYP (the latter in its pure definition and including both D2 and D3 dispersion corrections) with those provided by CASPT2 for activation energies and those provided by CCSD(T) for binding energies, using one and two water molecules plus the radicals as test systems, indicates that B3LYP-D3 underestimates the energy barriers, while BHLYP-based methods show a reasonably good performance. For the computations relative to the 18 and 33 water clusters, we adopted BHLYP-D3 as it has been found, in the test systems, to properly deal with both the radical/surface
binding and the radical-radical activation energies.\\
4. The morphology of the water cluster used for the simulations definitely affects the results of the computations. In particular, radicals would interact differently depending on
whether they sit on a cavity structure, where they can establish several weak inter-molecular interactions with the icy water molecules, in addition to the H-bond.\\
5. Taking into account the results described in points 1, 2 and 4, the mechanism that radical combination necessarily produces iCOMs is still to be validated, and should be taken with care in astrochemical models.

In order to make progresses, more accurate computations would be
needed, but they are not yet within the reach of the current computational
capacities. On the other hand, dynamical simulations would help to
understand the effect of the relative orientation of radicals upon
encounter.

\section{Supporting information}
Structures and errors of the benchmark study, PESs of water assisted H-transfer reactions on W18 and W33-cav, structures, binding energies and PESs (of radical-radical coupling, direct H-abstraction and water assisted H-transfer reactions) on the three analysed surfaces (W18, W33-side and W33-cav) at BHLYP-D2 and on W18 at B3LYP-D3, and optimized Cartesian coordinates of all the structures.

\begin{acknowledgement}
JER and CC acknowledge funding from the European Research Council (ERC) under the European Union's Horizon 2020 research and innovation program, for the Project ``the Dawn of Organic Chemistry" (DOC), grant agreement No 741002. AR is indebted to ``Ram{\'o}n y Cajal" program. MINECO (project CTQ2017-89132-P) and DIUE (project 2017SGR1323) are acknowledged. PU and NB acknowledge MIUR (Ministero dell’Istruzione, dell’Universit{\`a} e della Ricerca) and from Scuola Normale Superiore (project PRIN 2015, STARS in the CAOS - Simulation Tools for Astrochemical Reactivity and Spectroscopy in the Cyberinfrastructure for Astrochemical Organic Species, cod. 2015F59J3R). This project has received funding from the European Union’s Horizon 2020 research and innovation programme under the Marie Skłodowska-Curie grant agreement No 811312.\\
We thank Prof. Gretobape for fruitful and stimulating discussions.\\
Most of the calculations presented in this paper were performed using the GRICAD infrastructure (https://gricad.univ-grenoble-alpes.fr), which is partly supported by the Equip@Meso project (reference ANR-10-EQPX-29-01) of the programme Investissements d'Avenir supervised by the Agence Nationale pour la Recherche. Additionally this work was granted access to the HPC resources of IDRIS under the allocation 2019-A0060810797 attributed by GENCI (Grand Equipement National de Calcul Intensif).
\end{acknowledgement}

\bibliography{achemso-demo}

\newpage
\section{For TOC Only}
\includegraphics[width=9cm]{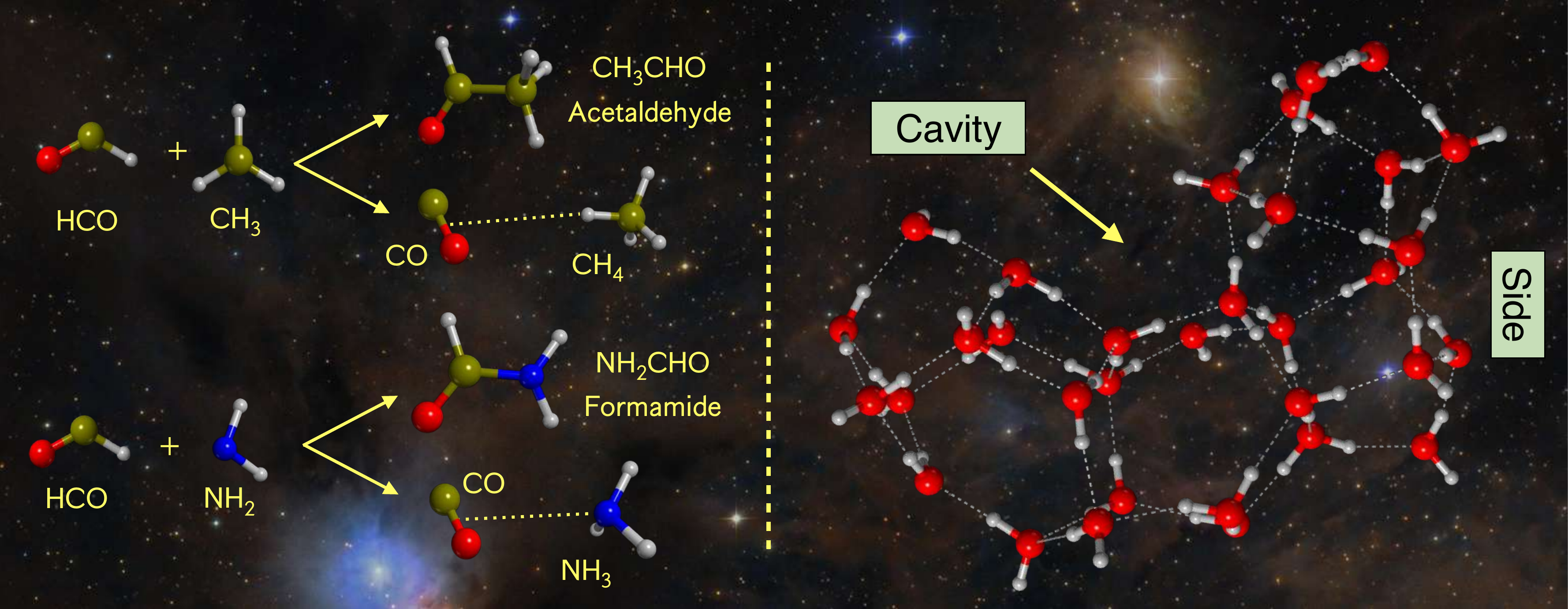}

\end{document}